\documentclass[a4paper]{article}
\usepackage[unicode]{hyperref}
\usepackage[dvipsnames,svgnames,x11names]{xcolor}
\usepackage[rmargin=3cm,lmargin=3cm,tmargin=3.5cm,bmargin=3.6cm]{geometry}
\usepackage{amsmath,amssymb}
\usepackage{lmodern}
\usepackage{iftex}
\ifPDFTeX
  \usepackage[T1]{fontenc}
  \usepackage[utf8]{inputenc}
  \usepackage{textcomp} 
\else 
  \usepackage{unicode-math}
  \defaultfontfeatures{Scale=MatchLowercase}
  \defaultfontfeatures[\rmfamily]{Ligatures=TeX,Scale=1}
\fi
\IfFileExists{upquote.sty}{\usepackage{upquote}}{}
\IfFileExists{microtype.sty}{
  \usepackage[]{microtype}
  \UseMicrotypeSet[protrusion]{basicmath} 
}{}
\makeatletter
\@ifundefined{KOMAClassName}{
  \IfFileExists{parskip.sty}{%
    \usepackage{parskip}
  }{
    \setlength{\parindent}{0pt}
    \setlength{\parskip}{6pt plus 2pt minus 1pt}}
}{
  \KOMAoptions{parskip=half}}
\makeatother
\usepackage{graphicx}
\makeatletter
\def\maxwidth{\ifdim\Gin@nat@width>\linewidth\linewidth\else\Gin@nat@width\fi}
\def\maxheight{\ifdim\Gin@nat@height>\textheight\textheight\else\Gin@nat@height\fi}
\makeatother
\setkeys{Gin}{width=\maxwidth,height=\maxheight,keepaspectratio}
\makeatletter
\def\fps@figure{htbp}
\makeatother
\newlength{\cslhangindent}
\newlength{\csllabelwidth}
\newlength{\cslentryspacingunit} 
\newenvironment{CSLReferences}[2] 
 {
  \setlength{\parindent}{0pt}
  \ifodd #1
  \let\oldpar\par
  \def\par{\hangindent=\cslhangindent\oldpar}
  \fi
  \setlength{\parskip}{#2\cslentryspacingunit}
 }%
 {}
\usepackage{calc}

\ifLuaTeX
\usepackage[bidi=basic]{babel}
\else
\usepackage[bidi=default]{babel}
\fi
\babelprovide[main,import]{american}

\def\languageshorthands#1{}
\ifLuaTeX
  \usepackage{selnolig}  
\fi
\IfFileExists{xurl.sty}{\usepackage{xurl}}{} 
\hypersetup{
  pdftitle={IGraph/M: graph theory and network analysis for
Mathematica},
  pdfauthor={Szabolcs Horvát, Jakub Podkalicki, Gábor Csárdi, Tamás
Nepusz, Vincent Traag, Fabio Zanini, Daniel Noom},
  pdflang={en-US},
  colorlinks=true,
  linkcolor={Maroon},
  filecolor={Maroon},
  citecolor={Blue},
  urlcolor={Blue},
  pdfcreator={LaTeX via pandoc}}

\title{IGraph/M: graph theory and network analysis \\ for Mathematica}

\usepackage[affil-it]{authblk}
\usepackage{orcidlink}
\author[1,2%
  \ensuremath\mathparagraph]{Szabolcs Horvát%
    \,\orcidlink{0000-0002-3100-523X}\,%
    }
\author[3%
  ]{Jakub Podkalicki%
    }
\author[4%
  ]{Gábor Csárdi%
    \,\orcidlink{0000-0001-7098-9676}\,%
    }
\author[5%
  ]{Tamás Nepusz%
    \,\orcidlink{0000-0002-1451-338X}\,%
    }
\author[6%
  ]{Vincent Traag%
    \,\orcidlink{0000-0003-3170-3879}\,%
    }
\author[7%
  ]{Fabio Zanini%
    \,\orcidlink{0000-0001-7097-8539}\,%
    }
\author[8%
  ]{Daniel Noom%
    }

\affil[1]{Max Planck Institute for Cell Biology and Genetics, Dresden,
Germany}
\affil[2]{Center for Systems Biology Dresden, Dresden, Germany}
\affil[3]{Independent Developer, Poland}
\affil[4]{RStudio}
\affil[5]{Independent Developer, Hungary}
\affil[6]{Centre for Science and Technology Studies, Leiden University,
Leiden, Netherlands}
\affil[7]{Lowy Cancer Research Centre, University of New South Wales,
Kensington, NSW, Australia}
\affil[8]{Independent Developer, Netherlands}
\affil[$\mathparagraph$]{Corresponding author}
\date{29 August 2022}

\begin{document}
\maketitle

\hypertarget{summary}{%
\section{Summary}\label{summary}}

IGraph/M\footnote{Available at
  \href{http://szhorvat.net/mathematica/IGraphM}{szhorvat.net/mathematica/IGraphM}
  and
  \href{https://github.com/szhorvat/IGraphM}{github.com/szhorvat/IGraphM}}
is an efficient general purpose graph theory and network analysis
package for Mathematica (\protect\hyperlink{ref-mathematica}{Wolfram
Research, Inc., 2022}). IGraph/M serves as the Wolfram Language
interfaces to the igraph C library
(\protect\hyperlink{ref-igraph}{Csárdi et al., 2022};
\protect\hyperlink{ref-Csardi2006}{Csárdi \& Nepusz, 2006}), and also
provides several unique pieces of functionality not yet present in
igraph, but made possible by combining its capabilities with
Mathematica's. The package is designed to support both graph theoretical
research as well as the analysis of large-scale empirical networks.

\hypertarget{statement-of-need}{%
\section{Statement of need}\label{statement-of-need}}

Mathematica contains
\href{https://reference.wolfram.com/language/guide/GraphsAndNetworks.html}{extensive
built-in functionality for working with graphs}. IGraph/M extends this
graph framework with many new functions not otherwise available in the
Wolfram Language, and also provides alternative and more featureful
open-source implementations of many of Mathematica's existing built-ins.
This makes it possible for Wolfram Language users to easily double-check
results, just as Python and R users can already do thanks to the
multiple different graph packages available in those languages. This is
particularly useful in graph theory where many results are just as
difficult to verify as to compute.

The only other independent graph theory package for Mathematica was
\href{https://reference.wolfram.com/language/Combinatorica/guide/CombinatoricaPackage}{Combinatorica}
(\protect\hyperlink{ref-combinatorica}{Pemmaraju \& Skiena, 2003}),
which has been mostly unmaintained since the introduction of the
\texttt{Graph} expression type with the release of Mathematica 8.0 in
2011. Despite this, not all of Combinatorica's functions have built-in
equivalents in Mathematica. IGraph/M provides replacements for almost
all of these old Combinatorica functions while offering much better
performance thanks to being implemented in a mix of C and C++ instead of
the Wolfram Language.

\hypertarget{design-goals-and-features}{%
\section{Design goals and features}\label{design-goals-and-features}}

One of the major appeals of Mathematica is its tightly integrated
nature: different functionality areas of the system can smoothly and
seamlessly interoperate with each other. In order to preserve this
benefit, a major design goal for IGraph/M was to integrate well into the
rest of the system. This is achieved by working directly with
Mathematica's native \texttt{Graph} data type, which is transparently
converted to igraph's internal format as needed. This makes IGraph/M
different from igraph's other high-level interfaces: igraph's internal
graph data structure is not exposed to users and the package does not
provide trivial operations which are already present in Mathematica,
such as adding or removing vertices and edges. Instead, priority is
given to functionality that delivers a true benefit over Mathematica's
own built-ins. While some of IGraph/M's functions may appear to simply
duplicate built-in functions, almost all of them provide additional
features not otherwise available in Mathematica. For example, unlike the
built-in \texttt{BetweenessCentrality{[}{]}} function,
\texttt{IGBetweenness{[}{]}} supports weighted graphs; unlike the
built-in \texttt{PageRankCentrality{[}{]}}, \texttt{IGPageRank{[}{]}}
takes into consideration self-loops and parallel edges; in contrast to
\texttt{IsomorphicGraphQ}, \texttt{IGIsomorphicQ} supports non-simple
graphs as well as vertex and edge colours; etc. In order to make it easy
to distinguish built-in symbols from those of IGraph/M, all functions in
this package have names starting with \texttt{IG}. The naming and
interface of functions is chosen so as to be familiar both to
Mathematica users and users of igraph's Python and R interfaces.

IGraph/M fully leverages the igraph C library's capabilities to
integrate into high-level host languages: Almost all computations are
interruptible in the usual manner. This feature is particularly
important for a graph theory package that provides multiple algorithms
with exponential or slower time complexity. Despite being implemented in
a compiled language, IGraph/M uses Mathematica's built-in random number
generator by default. Therefore, all its stochastic algorithms respect
seeds set with \texttt{SeedRandom{[}{]}} or \texttt{BlockRandom{[}{]}}
the same way as built-in functions would.

IGraph/M aims to exploit the interactive features of Mathematica
notebooks to improve user productivity. In this spirit, it includes an
interactive graph editor, \texttt{IGGraphEditor{[}{]}}, and supports
dynamically displaying the progress of many functions.

\hypertarget{use-cases-and-unique-features}{%
\section{Use cases and unique
features}\label{use-cases-and-unique-features}}

IGraph/M provides multiple unique features that are not present in the
core igraph library. Examples include exact graph colouring
(\protect\hyperlink{ref-VanGelder2008}{Van Gelder, 2008}), functions for
working with planar graphs and combinatorial embeddings, proximity graph
functions (\protect\hyperlink{ref-Kirkpatrick1985}{Kirkpatrick \& Radke,
1985}; \protect\hyperlink{ref-Toussaint1980}{Toussaint, 1980}) and aids
for working with spatial networks, functions for performing tests
related to the graph automorphism group, and several others. Some
features, such as \texttt{IGRealizeDegreeSequence}, are based on
original research of the authors
(\protect\hyperlink{ref-Horvat2021}{Horvát \& Modes, 2021}).
Additionally, there are many convenience functions that are helpful when
working with Mathematica's \texttt{Graph}, including a simplified system
for working with edge and vertex attributes based on the concept of
mapping functions over attribute values (\texttt{IGEdgeMap},
\texttt{IGVertexMap}).

\begin{figure}
\centering
\includegraphics[width=0.9\textwidth,height=\textheight]{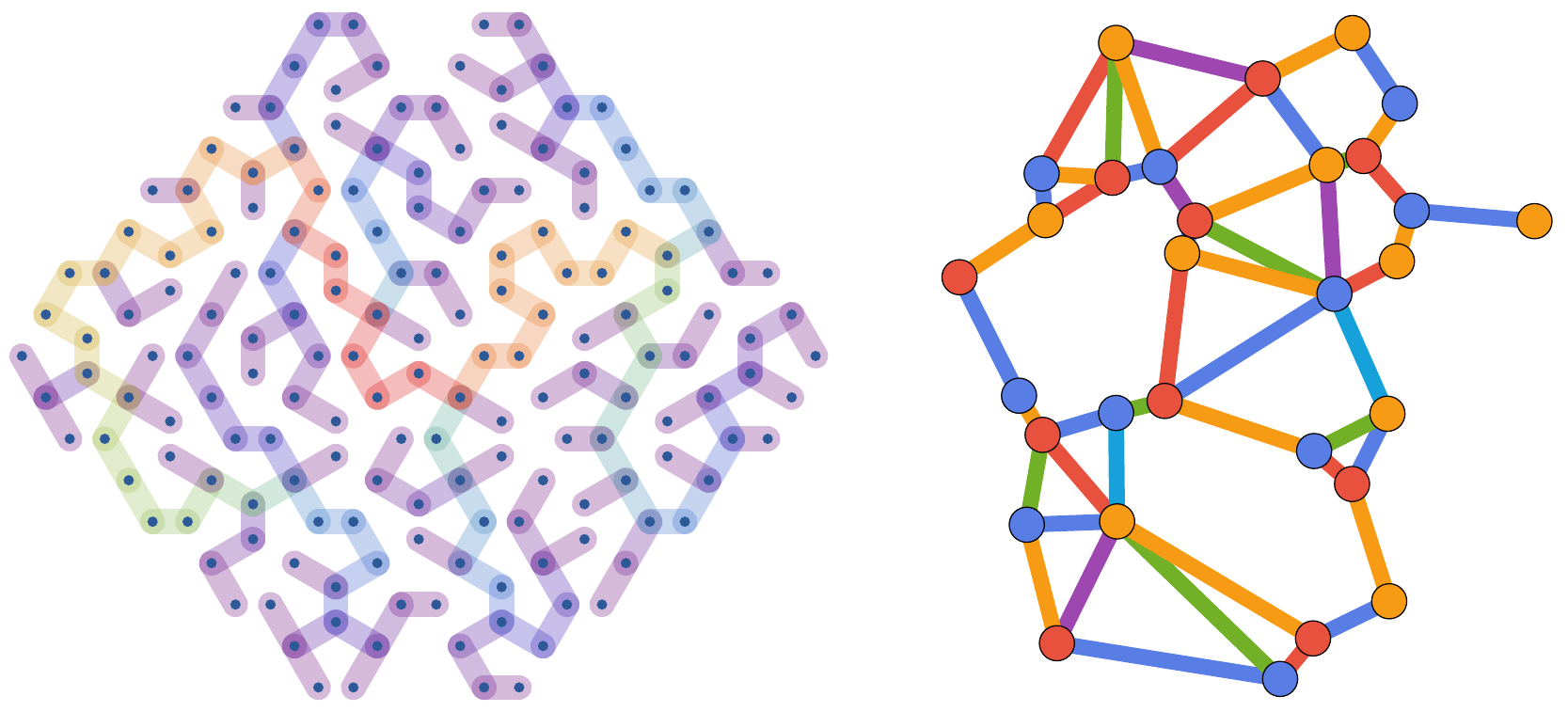}
\caption{Output of the above example code. \texttt{Out{[}2{]}} is on the
left, \texttt{Out{[}3{]}} on the right.\label{fig:fig1}}
\end{figure}

The following examples show a few of these features while also
demonstrating the concise idioms made possible by this package. The
output is shown in \autoref{fig:fig1}. Many more examples are found
\href{http://szhorvat.net/mathematica/IGDocumentation/}{in the package's
documentation}.

Load the package:

\begin{verbatim}
In[1]:= << IGraphM`
Out[1]= IGraph/M 0.6.2 (July 25, 2022)
        Evaluate IGDocumentation[] to get started.
\end{verbatim}

Create a random maze as a random spanning tree of a regular lattice
confined to a hexagon, and colour it according to betweenness
centrality:

\begin{verbatim}
In[2]:=
g = IGMeshGraph@IGLatticeMesh["CairoPentagonal", Polygon@CirclePoints[6, 6]];
t = IGRandomSpanningTree[g,
        VertexCoordinates -> GraphEmbedding[g], GraphStyle -> "ThickEdge"];
IGEdgeMap[ColorData["Rainbow"], EdgeStyle -> IGEdgeBetweenness /* Rescale, t]
\end{verbatim}

Compute and visualize a minimum vertex and edge colouring of the Gabriel
graph of a set of spatial points:

\begin{verbatim}
In[3]:=
IGGabrielGraph[RandomPoint[Disk[], 30],
    GraphStyle -> "Monochrome", VertexSize -> 1, EdgeStyle -> Thickness[1/40]
] //
  IGVertexMap[ColorData[106], VertexStyle -> IGMinimumVertexColoring] //
  IGEdgeMap[ColorData[106], EdgeStyle -> IGMinimumEdgeColoring]
\end{verbatim}

\hypertarget{implementation-notes}{%
\section{Implementation notes}\label{implementation-notes}}

IGraph/M is built using LTemplate
(\protect\hyperlink{ref-LTemplate}{Horvát, 2018}), an open-source system
that makes it easier to extend Mathematica using C++ code. IGraph/M also
serves as the primary driver of LTemplate development. Planar
graph-related functionality is implemented using the LEMON graph library
(\protect\hyperlink{ref-Dezso2011}{Dezső et al., 2011};
\protect\hyperlink{ref-lemon}{\hspace{0pt}Egerváry Research Group on
Combinatorial Optimization, 2014}). Spatial graph functions make use of
the nanoflann nearest neighbour search library
(\protect\hyperlink{ref-nanoflann}{Blanco \& Rai, 2022}).

\hypertarget{acknowledgements}{%
\section{Acknowledgements}\label{acknowledgements}}

This project has been made possible in part by grant number 2021-237461
from the Chan Zuckerberg Initiative DAF, an advised fund of Silicon
Valley Community Foundation. We thank Juho Lauri for feedback and ideas,
as well as for guidance with the implementation of the SAT-based exact
graph colouring functionality. Henrik Schumacher provided the basis for
the implementation of the mesh/graph conversion functions and some of
the proximity graph functions. Patrick Scheibe made available his
\href{http://wlplugin.halirutan.de/}{Wolfram Language plugin for
IntelliJ IDEA}, without which this project would have been much more
difficult to manage. Sz.~H.~is grateful to Carl Modes for ongoing
support and feedback regarding the paper, and to the
\href{https://www.wiko-berlin.de/}{Wissenchaftskolleg zu Berlin} for
time to work on spatial networks functionality in this package.

\hypertarget{references}{%
\section*{References}\label{references}}
\addcontentsline{toc}{section}{References}

\hypertarget{refs}{}
\begin{CSLReferences}{1}{0}
\leavevmode\vadjust pre{\hypertarget{ref-nanoflann}{}}%
Blanco, J. L., \& Rai, P. K. (2022). \emph{Nanoflann: A {C}++
header-only fork of {FLANN}, a library for nearest neighbor ({NN}) with
KD-trees} (Version 1.4.3) {[}Computer software{]}. GitHub.
\url{https://github.com/jlblancoc/nanoflann}

\leavevmode\vadjust pre{\hypertarget{ref-Csardi2006}{}}%
Csárdi, G., \& Nepusz, T. (2006). The igraph software package for
complex network research. \emph{InterJournal Complex Systems},
\emph{1695}(5), 1--9.
\url{http://www.interjournal.org/manuscript_abstract.php?361100992}

\leavevmode\vadjust pre{\hypertarget{ref-igraph}{}}%
Csárdi, G., Nepusz, T., Horvát, Sz., Traag, V., Zanini, F., \& Noom, D.
(2022). \emph{The igraph {C} library} (Version 0.9.9) {[}Computer
software{]}. Zenodo. \url{https://doi.org/10.5281/zenodo.3630268}

\leavevmode\vadjust pre{\hypertarget{ref-Dezso2011}{}}%
Dezső, B., Jüttner, A., \& Kovács, P. (2011). {LEMON} -- an open source
{C++} graph template library. \emph{Electron.~Notes
Theor.~Comput.~Sci.}, \emph{264}, 23--45.
\url{https://doi.org/10.1016/j.entcs.2011.06.003}

\leavevmode\vadjust pre{\hypertarget{ref-lemon}{}}%
\hspace{0pt}Egerváry Research Group on Combinatorial Optimization.
(2014). \emph{{LEMON} -- library for efficient modeling and optimization
in networks} (Version 1.3.1) {[}Computer software{]}.
\url{https://lemon.cs.elte.hu/}

\leavevmode\vadjust pre{\hypertarget{ref-LTemplate}{}}%
Horvát, Sz. (2018). \emph{LTemplate} (Version 0.5.4) {[}Computer
software{]}. GitHub. \url{https://github.com/szhorvat/LTemplate/}

\leavevmode\vadjust pre{\hypertarget{ref-Horvat2021}{}}%
Horvát, Sz., \& Modes, C. D. (2021). Connectedness matters: Construction
and exact random sampling of connected networks. \emph{Journal of
Physics: Complexity}, \emph{2}, 015008.
\url{https://doi.org/10.1088/2632-072X/abced5}

\leavevmode\vadjust pre{\hypertarget{ref-Kirkpatrick1985}{}}%
Kirkpatrick, D. G., \& Radke, J. D. (1985). A framework for
computational morphology. \emph{Machine Intelligence and Pattern
Recognition}, \emph{2}(C), 217--248.
\url{https://doi.org/10.1016/B978-0-444-87806-9.50013-X}

\leavevmode\vadjust pre{\hypertarget{ref-combinatorica}{}}%
Pemmaraju, S., \& Skiena, S. (2003). \emph{Computational discrete
mathematics: Combinatorics and graph theory with {Mathematica}}.
Cambridge University Press. \url{http://www.combinatorica.com}

\leavevmode\vadjust pre{\hypertarget{ref-Toussaint1980}{}}%
Toussaint, G. T. (1980). The relative neighbourhood graph of a finite
planar set. \emph{Pattern Recognition}, \emph{12}(4), 261--268.
\url{https://doi.org/10.1016/0031-3203(80)90066-7}

\leavevmode\vadjust pre{\hypertarget{ref-VanGelder2008}{}}%
Van Gelder, A. (2008). Another look at graph coloring via propositional
satisfiability. \emph{Discrete Applied Mathematics}, \emph{156}(2),
230--243. \url{https://doi.org/10.1016/j.dam.2006.07.016}

\leavevmode\vadjust pre{\hypertarget{ref-mathematica}{}}%
Wolfram Research, Inc. (2022). \emph{Mathematica} (Version 13.1)
{[}Computer software{]}. {Wolfram Research, Inc.}
\url{https://www.wolfram.com/mathematica/}

\end{CSLReferences}

\end{document}